\documentclass[sigconf, natbib=true]{acmart}
\usepackage{colortbl}
\usepackage{color}
\usepackage{booktabs}
\usepackage{hhline}
\usepackage{ulem} 
\usepackage{xspace}
\usepackage{url}
\usepackage{ulem}
\usepackage{subfigure}
\usepackage{enumitem}

\usepackage{color}
\definecolor{cGreen}{RGB}{0,150,0}
\definecolor{brown}{RGB}{139,64,0}

\newcommand{\ourname}{{DcRec}\xspace}

\AtBeginDocument{%
  \providecommand\BibTeX{{%
    \normalfont B\kern-0.5em{\scshape i\kern-0.25em b}\kern-0.8em\TeX}}}

\setcopyright{none}
\setcopyright{acmcopyright}
\copyrightyear{2022}
\acmYear{2022}

\acmPrice{15.00}

\setcopyright{acmcopyright}
\acmConference[CIKM '22]{Proceedings of the 31st ACM International Conference on Information and Knowledge Management}{October 17--21, 2022}{Atlanta, GA, USA}
\acmBooktitle{Proceedings of the 31st ACM International Conference on Information and Knowledge Management (CIKM '22), October 17--21, 2022, Atlanta, GA, USA}
\acmDOI{10.1145/3511808.3557583}
\acmISBN{978-1-4503-9236-5/22/10}

\begin{document}

\title{Disentangled Contrastive Learning for Social Recommendation}
\author{Jiahao Wu}
\affiliation{
    \institution{The Hong Kong Polytechnic University}
    \city{Hong Kong}
    \country{China}
}
\affiliation{
    \institution{Southern University of Science and Technology}
    \city{Shenzhen}
    \state{Guangdong}
    \country{China}
}
\email{jiahao.wu@connect.polyu.hk}

\author{Wenqi Fan}
\authornote{Corresponding Author.}
\affiliation{
    \institution{The Hong Kong Polytechnic University}
    \city{Hong Kong}
    \country{China}
}
\email{wenqifan03@gmail.com}

\author{Jingfan Chen}
\affiliation{
    \institution{The Hong Kong Polytechnic University}
    \city{Hong Kong}
    \country{China}
}
\affiliation{
    \institution{Centre for Artificial Intelligence and Robotics (HKISI\_CAS)}
    \city{Hong Kong}
    \country{China}
}
\email{jingfan.chen@connect.polyu.hk}

\author{Shengcai Liu}
\affiliation{
    \institution{Southern University of Science and Technology}
    \city{Shenzhen}
    \state{Guangdong}
    \country{China}
}
\email{liusc3@sustech.edu.cn}

\author{Qing Li}
\affiliation{
    \institution{The Hong Kong Polytechnic University}
    \city{Hong Kong}
    \country{China}
}
\email{csqli@comp.polyu.edu.hk}

\author{Ke Tang}
\affiliation{
    \institution{Southern University of Science and Technology}
    \city{Shenzhen}
    \state{Guangdong}
    \country{China}
}
\email{tangk3@sustech.edu.cn}

\hyphenpenalty=5000 
\tolerance=1000 
\begin{abstract}
Social recommendations utilize social relations to enhance the  representation learning for recommendations.Most social recommendation models unify   user representations for the user-item interactions (collaborative domain) and social relations (social domain). However, such an approach may fail to model the users'  heterogeneous behavior patterns in two domains, impairing the expressiveness of  user representations. In this work, to address such limitation, we propose a novel \textbf{D}isentangled \textbf{c}ontrastive learning framework for social \textbf{Rec}ommendations (\textbf{DcRec}). 
More specifically, we propose to learn disentangled users' representations from the item and social domains. 
Moreover, disentangled contrastive learning  is designed to perform knowledge transfer between disentangled users’ representations for  social recommendations.
Comprehensive experiments on various real-world datasets demonstrate the superiority of our proposed model. 
\end{abstract}



\maketitle

\section{Introduction}

\noindent As suggested by social correlation theories~\cite{socialInfluence2004},  users' preferences are likely to  be influenced by those around them. Based on this intuition, a bunch of works have been proposed to utilize the information of social relations in modeling users' preferences for items to  enhance recommendation performance in various online platforms (e.g., Facebook, WeChat, LinkedIn, etc.)~\cite{fan2019deep_daso, recsys2009learningTrustandDistrust, aaai2015trustsvd}, known as social recommendations~\cite{fan2020graph,fan2019graphRec, wu2019diffNet, wu2020diffNetPlus, fan2018deep}. 

\begin{figure}[t]
\centering
\vskip -0.10in
{\includegraphics[width=0.50\linewidth]{{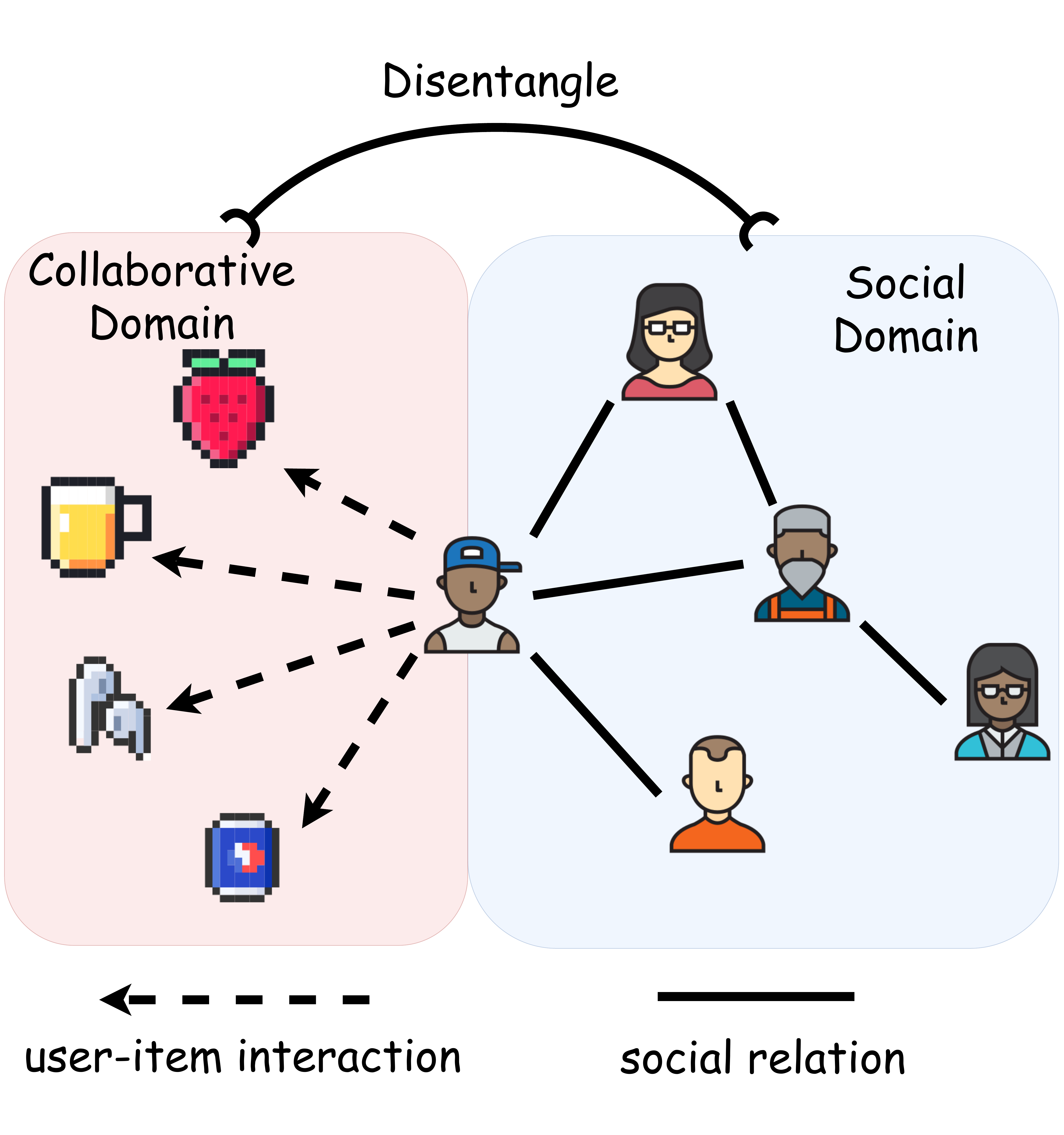}}}
\vskip -0.15in
\caption{Users are involved in both collaborative domain and social domain. The behavior patterns of users in two domains are semantically heterogeneous, where a user can have his/her personalized preferences towards items in the collaborative domain while connecting with social friends in the social domain.
}\label{fig:two-domains}
\vskip -0.25in
\end{figure}

In social recommendation, as shown in Figure~\ref{fig:two-domains}, users are interacted with different objectives (i.e., items and social friends) with distinct purposes in each domain (i.e., collaborative domain and social domain)~\cite{fan2019deep_daso}.
Therefore, the behavior patterns of users in two domains can be heterogeneous.
In a real scenario, a user tends to connect with friends of his/her friends but he/she are not likely to purchase the items serving similar functions in a short period of time. However, off-the-shelf manners on modeling users' preferences adopt unified users' representations for user-item interactions and user-user social relationships. For instance, Diffnet++~\cite{wu2019diffNet} utilizes a node-level attention layer to fuse the users' representations from social domain and collaborative domain. DSCF~\cite{fan2019dscf} uses unified users' representations to capture social information and user-item interaction information. 
They are insufficient to model users' heterogeneous behavior patterns towards social friends and items in social recommendations, which may undermine the representation learning in each domain. 


To address this problem, in this work,  we propose to \textit{disentangle} user behaviors into two domains, so as to learn disentangled users' representations in the social recommendations task, instead of the unified users' representations. The main challenge is how to learn such disentangled users' representations in two domains while transferring knowledge from social domain to collaborative domain for social recommendations. 

Recently,  Self-Supervised Learning (SSL), a prevalent novel learning paradigm, has been proven beneficial to the tasks in a wide range of fields~\cite{simclr_v1, nlp2018contra, 2018contraCoding, kaiming2020moco, comprehensive2021sslGraphSurvey, DGCL2021Li}. The main idea of SSL is to utilize  self-supervision signals from unlabeled data by maximizing the mutual information between different augmented views of the same instance (i.g., user or item)~\cite{alignUniform2020ICMLcontra}, in the representation learning process of which SSL increases the informativeness of those views and enables knowledge transferring between those views~\cite{yu2021MHCN}.
For instance, SGL~\cite{he2021sgl} devises various auxiliary tasks to learn the structure relatedness knowledge between different views and utilize it to supplement supervision signals. MHCN~\cite{yu2021MHCN} constructs different hyper-graphs and maximizes the mutual information between the users and the hyper-graphs to learn the hierarchical structure information and transfer it into the learned nodes' representations.

Motivated by the advantage of SSL in transferring knowledge, we develop a novel contrastive learning-based framework to solve the aforementioned issue.
More specifically, domain disentangling is introduced to disentangle users' behaviors into  collaborative domain and social domain. 
Moreover, we propose disentangled contrastive learning objectives to transfer knowledge from social domain to collaborative domain by maximizing the mutual information between  disentangled representations. Notably, while DGCL~\cite{DGCL2021Li} proposes an implicitly disentangled contrastive learning method to capture multiple aspects of graphs, we explicitly disentangle the data from different domains and propagate the representations of nodes based on independent adjacency matrices.
The main contributions of this paper can be summarized as follows:
\begin{itemize} [leftmargin=*]
    \item We introduce a principled approach to learn users' representations, where disentangled users' representations can be learned to reflect their preferences towards items and social friends in two domains.  
    
    
    \item We propose  a novel \textbf{D}isentangled \textbf{c}ontrastive learning framework for social \textbf{Rec}ommendations ({\textbf{\ourname}}), which can harness the power of contrastive learning to transfer knowledge from social domain to collaborative domain.
    
    
    
    \item We conduct comprehensive experiments on various real-world datasets to show the superiority of the proposed model and study the  effectiveness of each module.
\end{itemize}

\section{Methodology}
\label{sec:methodology}
We first introduce definitions and notations used in this paper.
We utilize $\mathcal{U}$ to stand for the user set and $\mathcal{I}$ to stand for the item set. Let $m =|\mathcal{U}|$ defines the number of users and $n=|\mathcal{I}|$ denotes the number of items. 
Then, we denote the user-item interactions matrix in the collaborative domain by $\mathbf{A}_I\in \mathbb{R}^{m\times n}$ and social relations matrix in the social domain by $\mathbf{A}_S\in \mathbb{R}^{m\times m}$.
In addition, we  use dense vectors to represent users and items (i.e., embeddings), where $\textbf{P}_S\in \mathbb{R}^{m\times d}$ and $\textbf{P}_I\in \mathbb{R}^{m\times d}$   denote users' embeddings with $d$ dimension in social domain and collaborative domain, respectively. $\textbf{Q}_I \in \mathbb{R}^{n\times d}$   denotes  items' embeddings in the collaborative domain.

\subsection{An Overview of the Proposed Framework}
\noindent 
In this work, we propose a \textbf{D}isentangled \textbf{c}ontrastive learning framework for social \textbf{Rec}ommendations ({\textbf{\ourname}}), which follows the general paradigm of self-supervised contrastive learning via maximizing the representation agreement between different views on the same instance~\cite{he2021sgl,simclr_v1,jin2020sslGraphSurvey}.

\begin{figure*}[htbp]
\centering
\vskip -0.18in
{\includegraphics[width=0.8\linewidth]{{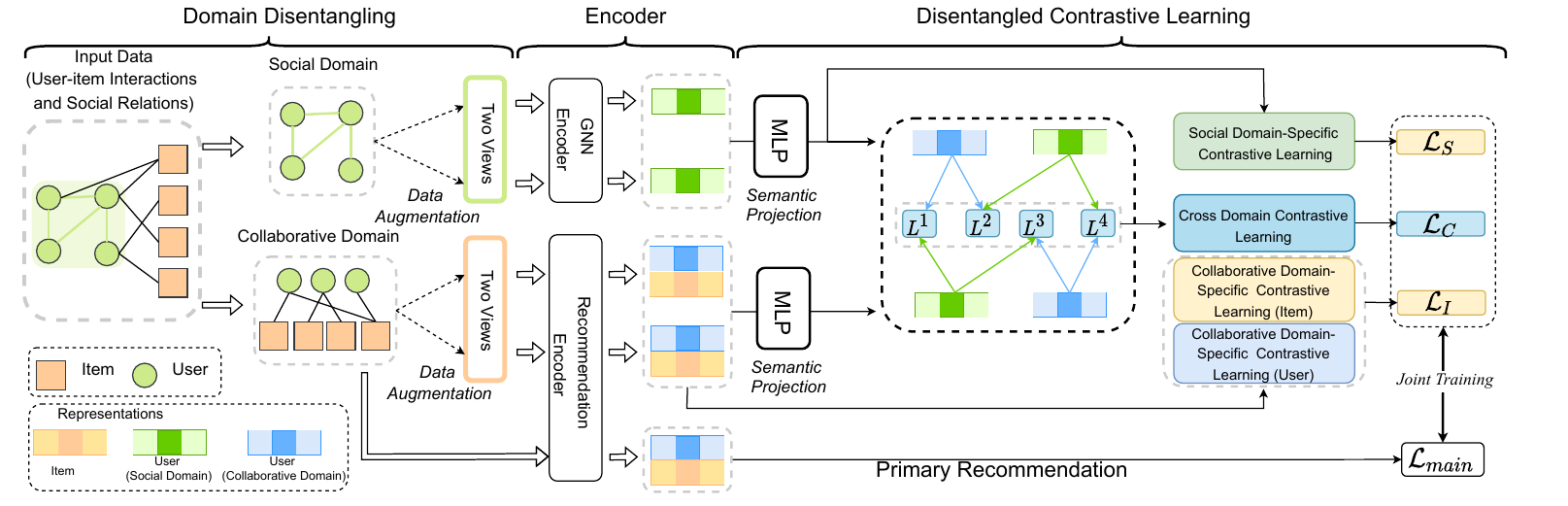}}}
\vskip -0.15in
\caption{The overall architecture of the proposed disentangled contrastive learning
 for social recommendations (DcRec).}
\vskip -0.18in

\label{fig:entire_frame}
\end{figure*}

The architecture of the proposed model is shown in Figure ~\ref{fig:entire_frame}.
More specifically, the proposed architecture consists of three main components: (1) \textbf{Domain Disentangling}, which is devised to disentangle the input data into two sub-domains; (2) \textbf{Encoder}, where we use different encoders on two domains to learn representations from two different views;  (3) \textbf{Disentangled Contrastive Learning}, which aims to transfer the knowledge from the social domain into recommendation modeling task by jointly optimizing the disentangled contrastive learning tasks and main recommendation task. 

\subsection{Domain Disentangling}

To mitigate the influence caused by the semantic discrepancy between social domain and collaborative domain, we disentangle the input data into two domains, which will be represented by a user-item interactions matrix $\mathbf{A}_I$ in the collaborative domain and social relations matrix $\mathbf{A}_S$ in the social domain, respectively.

After domain disentanglement, we perform  data augmentation  to obtain different views for the data in each domain. 
Since the data in social recommendations can be naturally represented as graph~\cite{fan2019graphRec}, the inputs (i.e., user-item interactions $\mathbf{A}_I$  and social relations $\mathbf{A}_S$) can be augmented via graph-based data augmentation methods~\cite{nips2020graphcl,jin2020sslGraphSurvey}, such as  \textit{Edge Adding}, \textit{Edge Dropout}, and \textit{Node Dropout}, which  can be formulated as follows:

\centerline{
    $\mathbf{A}_S^{(1)} = H_S^{(1)}(\mathbf{A}_S),\mathbf{A}_S^{(2)} = H_S^{(2)}(\mathbf{A}_S),$
}
\centerline{
    $\mathbf{A}_I^{(1)} = H_I^{(1)}(\mathbf{A}_I),\mathbf{A}_I^{(2)} = H_I^{(2)}(\mathbf{A}_I),$
}

\noindent where $H_S^{(\cdot)}$ and $H_I^{(\cdot)}$ denote the independent augmentation functions to generate two views in social domain and collaborative domain, respectively. 

\subsection{Encoder} 
\label{sub-sec:encoder}
To model the user-item interactions and social relations, we utilize encoders to learn representations for users and items in each domain. Furthermore, to ensure semantic consistency while conducting cross-domain contrastive learning between users' representations from two different domains, we also project the users' representations into the same semantic space. 
Here, we use  $Rec(\cdot)$ and $F(\cdot)$ to represent the encoder in the item and social domains, respectively. 
Note that any Collaborative Filtering (CF) based models (\textit{e.g.}, MF~\cite{Koren2009MF}, NeuMF~\cite{www2017NeuMF}, and LightGCN~\cite{he2020lightgcn}.) can be set as the recommendation encoder  $Rec(\cdot)$ in the collaborative domain, and we can set Graph Neural Networks (GNNs) methods~\cite{fan2022graph,derr2020epidemic,fan2021jointly} as social encoder $F(\cdot)$.

\subsubsection{\textbf{Recommendation Encoder in Collaborative Domain}.}
The collaborative domain encoder aims to learn the representations of users and items by encoding the user-item interactions (i.e., $\mathbf{A}^{(1)}_I$ and $\mathbf{A}^{(2)}_I$) from two augmented views. 
In practice, a simple yet effective GNN-based recommendation model LightGCN~\cite{he2020lightgcn} is used as collaborative domain encoder $Rec(\cdot)$ and
we can obtain users' and items' representations from two views (i.e., $\mathbf{A}^{(1)}_I$ and $\mathbf{A}^{(2)}_I$) as:
\begin{align}
    \textbf{U}^{(1)}_{I},\textbf{V}^{(1)}_{I}= \text{Rec}(\mathbf{A}_I^{(1)};&\Theta_I),
    \textbf{U}^{(2)}_{I},\textbf{V}^{(2)}_{I}= \text{Rec}(\mathbf{A}_I^{(2)};\Theta_I), 
\end{align}
where  $\Theta_I$ is the parameter of the recommendation encoder. $\textbf{U}_I^{(1)}\in \mathbb{R}^{m\times d}$, $\textbf{U}_I^{(2)}\in \mathbb{R}^{m\times d}$, $\textbf{V}_I^{(1)}\in \mathbb{R}^{n\times d}$ and $\textbf{V}_I^{(2)}\in \mathbb{R}^{n\times d}$ are learned representations of users and items from two views for contrastive learning. 
Furthermore, we also use this encoder $Rec(\cdot)$ to train our primary task via BPR loss~\cite{he2020lightgcn} (Eq.~\ref{eq:bpr_prediction})  and learn final users and items representations (i.e., $\textbf{U}\in \mathbb{R}^{m\times d}$ and $\textbf{V}\in \mathbb{R}^{n\times d}$) on user-item interactions (i.e., $\mathbf{A}_I$) for making  predictions  as follows:

\centerline{
    $    \textbf{U},\textbf{V}= \text{Rec}(\mathbf{A}_I;\Theta_I).$
}

\subsubsection{\textbf{GNNs Encoder in Social Domain}.}
The   encoder in social domain aims at learning users' representations by capturing social relations among users. Here, due to the excellent expressiveness of GNNs in modeling graph structural data~\cite{fan2019graphRec}, we adopt a general GNNs method~\cite{kipf2016gcn} as an encoder (i.e., $F(\cdot)$) to obtain user representations in social domain as follows:
\begin{align}
    \textbf{U}^{(1)}_{S}= \text{F}(\mathbf{A}_S^{(1)};\Theta_S), \textbf{U}^{(2)}_{S}= \text{F}(\mathbf{A}_S^{(2)};\Theta_S),
\end{align}
where $\boldsymbol{\Theta}_S$ denotes the parameters of GNNs encoder in the social domain. $\textbf{U}_S^{(1)}\in \mathbb{R}^{m\times d}$ and $\textbf{U}_S^{(2)}\in \mathbb{R}^{m\times d}$ are learned representations of users from two views for contrastive learning in social domain.

\subsubsection{\textbf{Semantic Projection}} Since users' representations learned from collaborative domain and social domain are semantically heterogeneous, we propose to  project them into the same semantic space. Specifically, in social domain, we adopt Multilayer Perceptrons (\textbf{MLPs}) to perform such projection on users' representations  as:
 $\widetilde{\textbf{U}}^{(1)}_S = \text{MLP}(\textbf{U}^{(1)}_S; \theta_S),\widetilde{\textbf{U}}^{(2)}_S = \text{MLP}(\textbf{U}^{(2)}_S; \theta_S),$
where $\theta_S$  is the set of MLPs' parameters in the social domain.  Analogously, we can obtain users' representations $\widetilde{\textbf{U}}^{(1)}_I$ and $\widetilde{\textbf{U}}^{(2)}_I$ in the collaborative domain via MLP with parameters $\theta_I$. 

\subsection{Disentangled Contrastive Learning}
Disentangled contrastive learning consists of cross-domain contrastive learning and domain-specific contrastive learning. We devise cross-domain contrastive learning so as to transfer the knowledge from socia domain to collaborative domain.
In order to take advantage of self-supervision signals from unlabeled data under contrastive learning paradigm~\cite{he2021sgl,simclr_v1,jin2020sslGraphSurvey},  we introduce domain-specific loss to maximize the representation agreement between different views on the same instance in each domain.
\subsubsection{\textbf{Cross-domain Contrastive Learning Loss.}} In order to transfer the knowledge from social domain to collaborative domain, we design cross-domain contrastive learning loss  based on the projected users' representations (i.e., $\widetilde{\textbf{U}}^{(1)}_S, \widetilde{\textbf{U}}^{(2)}_S, \widetilde{\textbf{U}}^{(1)}_I, \widetilde{\textbf{U}}^{(2)}_I$) as follows:
\begin{equation}
\label{eq:cross-domain}
    \mathcal{L}_C = 
    L(\widetilde{\textbf{U}}^{(1)}_S,\widetilde{\textbf{U}}^{(1)}_I)+ L(\widetilde{\textbf{U}}^{(1)}_S,\widetilde{\textbf{U}}^{(2)}_I)+ L(\widetilde{\textbf{U}}^{(2)}_S,\widetilde{\textbf{U}}^{(1)}_I)+ L(\widetilde{\textbf{U}}^{(2)}_S,\widetilde{\textbf{U}}^{(2)}_I),\; 
\end{equation}
where $L(\cdot,\cdot)$ denotes a common contrastive learning loss which distinguishes the representations of the same users in these different views from other users' representations~\cite{www2021adaptive, Prototype2021GCL, he2021sgl}, where $L(\widetilde{\textbf{U}}^{(1)}_S,\widetilde{\textbf{U}}^{(1)}_I)$, $L(\widetilde{\textbf{U}}^{(1)}_S,\widetilde{\textbf{U}}^{(2)}_I)$, $L(\widetilde{\textbf{U}}^{(2)}_S,\widetilde{\textbf{U}}^{(1)}_I)$ and $L(\widetilde{\textbf{U}}^{(2)}_S,\widetilde{\textbf{U}}^{(2)}_I)$ are denoted by $L^{2}$, $L^{3}$, $L^{1}$ and $L^{4}$ in Figure~\ref{fig:entire_frame}, respectively.

Due to  symmetric property in  two contrasted views, a common contrastive learning loss  $L(\cdot,\cdot)$  can be formally given by:
\begin{align}
    L(\textbf{Z}^{(1)},\textbf{Z}^{(2)}) = \frac{1}{2w}\sum\limits_{j=1}^{w}\left[loss(\textbf{z}_j^{(1)},\textbf{z}_j^{(2)})+loss(\textbf{z}_j^{(2)},\textbf{z}_j^{(1)})\right],
\end{align}
where $\textbf{Z}^{(1)} \in \mathbb{R}^{w\times d}$ and $\textbf{Z}^{(2)} \in \mathbb{R}^{w\times d}$ are instances' representations in two different views, and $\textbf{z}^{(1)}_j$ and $\textbf{z}^{(2)}_j$ are corresponding representations of $u$-th instance (i.e., users and items) in two views and $w \in \{m,n\}$.
Inspired by the design in~\cite{www2021adaptive}, the $loss(\textbf{z}_j^{(1)},\textbf{z}_j^{(2)}) $ can be formulated as:
\begin{equation}
\label{eq:ssl_loss_instance}
    loss(\textbf{z}_j^{(1)},\textbf{z}_j^{(2)}) = -\log{\frac{e^{\Psi(\textbf{z}_j^{(1)},\textbf{z}_j^{(2)})}}{e^{\Psi(\textbf{z}_j^{(1)},\textbf{z}_j^{(2)})}+\sum\limits_{v\in \{1,2\}}\sum\limits_{k\neq j}{e^{\Psi(\textbf{z}_j^{(1)},\textbf{z}_k^{(v)})}} }
    },
\end{equation}
where $\Psi(\textbf{z}_1,\textbf{z}_2) = s(\textbf{z}_1,\textbf{z}_2)/\tau$, measuring the cosine similarity between two representations, 
and $\tau$ is the temperature parameter. 

\subsubsection{\textbf{Domain-specific Contrastive Learning Loss}.} To enhance the expressiveness of the learned representations for each instance in each domain, we design domain-specific contrastive learning loss in the two domains:
\begin{align}
    \text{Collaborative Domain: }&  \mathcal{L}_I = L(\textbf{U}^{(1)}_I, \textbf{U}^{(2)}_I) + L(\textbf{V}^{(1)}_I, \textbf{V}^{(2)}_I), \\
    \text{Social Domain: }& \mathcal{L}_{S} = L(\widetilde{\textbf{U}}^{(1)}_S, \widetilde{\textbf{U}}^{(2)}_S), 
\end{align}
where $\mathcal{L}_{S}$ and $\mathcal{L}_{I}$  are domain-specific contrastive learning losses for the social domain and collaborative domain, respectively. 
Note that in practice, the projected representations in the social domain are used to conduct the domain-specific contrastive learning due to the promising results in our experiments. 

\subsection{Model Optimization}
\subsubsection{\textbf{Primary Recommendation Task}} Given the learned representation $\textbf{u}_u$ and $\textbf{v}_i$ for user $u$ and item $i$, we adopt a widely used inner product to predict the score for measuring how likely the user $u$ will interact with item $i$ as: $\hat{y}_{ui} = \textbf{u}^T_u\textbf{v}_i$.

To optimize the primary recommendation task,  we choose Bayesian Personalized Ranking (BPR) loss~\cite{rendle2009bpr}, formulated as:
\begin{equation}
\label{eq:bpr_prediction}
    \mathcal{L}_{main} = \sum\limits_{(u,i,j)\in \mathcal{O}}- \log{\sigma(\hat{y}_{ui}- \hat{y}_{uj})
    },
\end{equation}
where $\mathcal{O}=\{(u,i,j)|(u,i)\in\mathcal{O}^+,(u,j)\in\mathcal{O}^-\}$. Here, $\mathcal{O^+}$ is the set of observed interactions and $\mathcal{O^-}$ is the set of unobserved ones. 
\subsubsection{\textbf{Joint Training}}
To improve the recommendation performance by our proposed model with  disentangled contrastive learning, we adopt a joint training strategy to optimize both the recommendation loss and contrastive learning loss:

\centerline{
    $\mathcal{L} = \mathcal{L}_{main} + \lambda_1(\mathcal{L}_{I}+\mathcal{L}_{S}) + \lambda_2\mathcal{L}_{C} + \lambda_3\Vert \zeta \Vert_2,$
}
\noindent where    $\lambda_1$, $\lambda_2$, and $\lambda_3$ are hyperparameters to balance the contributions of various contrastive learning losses and regularization. 

\section{Experiments}
\label{sec:experiments}

Due to the space limitation, the detailed experimental settings, evaluation metrics, baseline and implementations could be found in Appendix Section.

\subsection{Experiment Results}
\subsubsection{\textbf{Overall Performance Comparison}}

In this part, we verify the effectiveness of  the proposed model {\ourname} for recommendation performance. The overall comparisons between our model with other methods are given in Table~\ref{tab:performance_comparison}. The baselines used for comparisons range from classical MF-based models, GNNs-based models to recent SSL enhanced models. According to the results, we can draw the following findings:

\begin{itemize}[leftmargin=*]
    
    \item The proposed {\ourname} achieves the best recommendations performance under all evaluation metrics on Dianping and Ciao datasets, achieving promising improvement over the strongest baselines that are marked with underlines. Compared to SSL-enhanced social recommendation baselines
    , our method incorporates advanced components to learn representations from social and collaborative domains by disentangled contrastive learning. 
    
    \item In most cases, SSL-enhanced methods outperform those methods without SSL on Dianping and Ciao datasets over various metrics. 
    
    This observation demonstrates the effectiveness of SSL for recommendations, which empirically shows the necessity of designing the domain-specific contarstive learning part.
    
\end{itemize}
\begin{table}[t]
\centering
\caption{Overall Performance Comparison}
\vskip -0.12in
\label{tab:performance_comparison}
\scalebox{0.75}{
\begin{tabular}{c|ccc|ccc} 
\hline
\textbf{Dataset} & \multicolumn{3}{c|}{Dianping} & \multicolumn{3}{c}{Ciao}                                                                                                         \\ 
\hline
Metrics          & NDCG    & Recall  & Precision & NDCG                     & Recall                   & Precision                            \\
\hline\hline

BPR~\cite{rendle2009bpr}              & 0.0188 & 0.0189 & 0.0145   & 0.0226                  & 0.0155                  & 0.0177                 \\
SBPR~\cite{cikm2014sbpr}             & 0.0162 & 0.0203 & 0.0217   & 0.0284                  & 0.0191                  & 0.0223                      \\
SoRec~\cite{cikm2008SoRec}            & 0.0149 & 0.0149 & 0.0111   & 0.0220                  & 0.0147                  & 0.0165                \\
SocialMF~\cite{recSys2010SocialMF}         & 0.0144 & 0.0148 & 0.0109   & 0.0218                  & 0.0154                  & 0.0166                \\
\hline
Diffnet~\cite{wu2019diffNet}          & 0.0241 & 0.0242 & 0.0181   & 0.0280 & 0.0182 & 0.0213\\
NGCF~\cite{wang2019ngcf}             & 0.0287 & 0.0289 & 0.0221   & 0.0232 & 0.0157 & 0.0183 \\
LightGCN~\cite{he2020lightgcn}         & 0.0396  & 0.0384  & 0.0292    & 0.0326                   & \underline{0.0224}                   & 0.0253  \\
LightGCN+Social       & 0.0392  & 0.0381  & 0.0289    & 0.0326                   & 0.0220                   & 0.0254  \\
\hline
SGL~\cite{he2021sgl}              & \underline{0.0421}  & \underline{0.0414}  & \underline{0.0310}    & \underline{0.0340}  & 0.0223  & \underline{0.0266} \\
SGL+Social            & 0.0405  & 0.0402  & 0.0299    & 0.0334  & 0.0221  & \underline{0.0266}   \\
MHCN~\cite{yu2021MHCN}             & 0.0398 & 0.0399 & 0.0296   & 0.0315                  & 0.0218                  & 0.0252 \\
SEPT~\cite{yu2021socially}             & 0.0378 & 0.0371 & 0.0286   & 0.0313                  & 0.0212                  & 0.0250\\
\textbf{{\ourname}({w/o~MLP})}    &   0.0417  &  0.0409 & 0.0307  &  0.0338   &    0.0217                &           0.0263         \\ 
\textbf{{\ourname}}       & \textbf{0.0443}  & \textbf{0.0441}  & \textbf{0.0327}    & \textbf{0.0366}                   & \textbf{0.0243}                   & \textbf{0.0280} \\ 
\hline\hline

\%Improv.       & 5.31\%  & 6.63\%  & 5.29\%    &   6.46\%                       &    8.39\%        &           5.32\% \\
\hline
\end{tabular}
}
\vskip -0.12in
\end{table}

\begin{table}[t]
\centering
\caption{Ablation study of $\lambda_1$ and $\lambda_2$ on dataset Ciao}
\vskip -0.12in
\label{tab:ablation_lambda}
\scalebox{0.6355}{
\begin{tabular}{ll|lll|ll|lll}
\hline
\checkmark&                                 & \multicolumn{3}{l|}{Metrics}                           &  &\checkmark                                     & \multicolumn{3}{l}{Metrics}   \\
\hline
$\lambda_1$ & $\lambda_2$ & NDCG             & Recall           & Precision        & $\lambda_1$   & $\lambda_2$   & NDCG    & Recall  & Precision \\ \hline\hline
0                         & 0.01                      & 0.03488          & 0.02436          & 0.02668          & 0.01 & 0                           & 0.03617 & 0.02394 & 0.02794   \\
0.001                     & 0.01                      & 0.03498          & 0.02416          & 0.02662          & \textbf{0.01} & \textbf{0.001}                       & \textbf{0.03657} & \textbf{0.02430} & \textbf{0.02795}   \\ 
\textbf{0.01}             & \textbf{0.01}             & \textbf{0.03629} & \textbf{0.02405} & \textbf{0.02773} &  0.01& 0.01 & 0.03629 & 0.02405 & 0.02773   \\
0.1                       & 0.01                      & 0.02735          & 0.01760          & 0.02191          & 0.01 & 0.1                         & 0.03635 & 0.02412 & 0.02803   \\
1                         & 0.01                      & 0.00814          & 0.00513          & 0.00681          & 0.01 & 1                           & 0.03618 & 0.02398 & 0.02803   \\
10                        & 0.01                      & 0.00103          & 0.00088          & 0.00075          & 0.01 & 10                          & 0.03626 & 0.02419 & 0.02803   \\

\hline
\end{tabular}
}
\vskip -0.2in
\end{table}

\subsubsection{\textbf{Ablation Study on MLP}} As shown in Table~\ref{tab:performance_comparison}, without the MLP for semantic projection, the performance degrades on all metrics for two datasets. This empirically verifies the effectiveness of the design of semantic projection.

\subsubsection{\textbf{Sensitivity Study on $\lambda_1$ and $\lambda_2$}} As shown in Table~\ref{tab:ablation_lambda}, we present the performance of {\ourname} with different values of $\lambda_1$ and $\lambda_2$. 
From the left part, studying the $\lambda_1$, we can find that setting a large value could lead to great drop on performance. We argue that this is because a large weight for domain-specific contrastive learning can lead the optimization towards instance discrimination (the objective of SSL) instead of the recommendation objective. From the right part, we found a similar phenomenon, but the best value for $\lambda_2$ is much smaller. We justify that this is because the proper amount of social information is beneficial to recommendation task but too much is harmful.

\section{Conclusion}
\label{sec:conclusion}

To model the heterogeneous behavior patterns of users in social domain and item domain, we proposed a disentangled contrastive learning framework for social recommendations, which disentangles two domains and learns users' representations in each domain, respectively. Furthermore, we devised cross-domain contrastive learning to transfer knowledge from the social domain to the item domain, so as to enhance the recommendation performance. The extensive experiments conducted on two real-world datasets demonstrate the effectiveness of our proposed model and its superiority over previous works.

\section*{ACKNOWLEDGMENTS}
\label{sec:ACKNOWLEDGMENTS}
This work was supported in part by the Guangdong Provincial Key Laboratory under Grant 2020B121201001, in part by the Program for Guangdong Introducing Innovative and Entrepreneurial Teams under Grant 2017ZT07X386, and in part by the Shenzhen Peacock Plan under Grant KQTD2016112514355531.
Qing LI, Wenqi FAN, and Jingfan CHEN are partly supported by NSFC (project no. 62102335), an internal research fund from The Hong Kong Polytechnic University (project no. P0036200 and P0042693), a General Research Fund from the Hong Kong Research Grants Council (Project No.: PolyU 15200021 and PolyU 15207322).

\normalem

\newpage
\clearpage
\balance
\bibliographystyle{ACM-Reference-Format}
\bibliography{references/references}

\newpage

\section{Appendix}
\label{sec:appendix}

\subsection{Related Work}
\noindent In this section, we will review previous work on social recommendation and SSL in recommendation. 

\subsubsection{GNNs-based Social Recommendations}
Recommender systems play an increasing role in various applications in our daily life~\cite{chen2022knowledge,fan2021attacking,zhao2021autoloss,zhaok2021autoemb}, such as online shopping and social media. 
It's a folklore knowledge that users' preferences and decisions are swayed by those who are socially connected with them~\cite{mcpherson2001birds,marsden1993network, liu2020improving}. Therefore, it's intuitive to integrate the social connection information into recommendation modeling. 
Graph neural networks (GNNs) are emerging in recent years and have achieved promising performances in many tasks~\cite{liu2021trustworthy}, as well as recommendation~\cite{fan2020graph}. The intuition of those prevalent GNN-based recommendation models (e.g., NGCF~\cite{wang2019ngcf} and LightGCN~\cite{he2020lightgcn}) is to utilize GNN to aggregate neighbors' information so as to learn the supervisory signals of high-order structure. To be more specific, in the field of social recommendation, GNNs also gained considerable attention. GraphRec~\cite{fan2019graphRec} is the very first GNN-based social recommendation method by modeling user-item interactions and social connections as graphs. DANSER~\cite{wu2019DANSER} proposes dual graph attention networks to collaboratively learn representations for two-fold social effects. DiffNet~\cite{wu2019diffNet} models the recursive dynamic social diffusion in social recommendation with a layer-wise propagation structure. DASO~\cite{fan2019deep_daso} adopts a bidirectional mapping method to transfer users' information between social domain and item domain using adversarial learning.

\subsubsection{Self-Supervised Learning in Recommendations}
Self-supervised learning (SSL) is becoming more and more prevalent in recent years, which is utilized to extract supervisory signals from data to learn expressive representations without labels. SSL was first applied in tasks on image data~\cite{simclr_v1, simclr_v2, kaiming2020moco}, gaining promising performance. Naturally, there are tons of work extending the SSL to graph domain~\cite{jin2020sslGraphSurvey, comprehensive2021sslGraphSurvey, liu2022self}. 
Since SSL has gained promising performance on graph representation learning, some recent studies also verify that this paradigm also work in the scenario of recommendation. SGL~\cite{he2021sgl} proposes to devise an auxiliary self-supervised task to supplement the classical supervised task, mitigating the problem of imbalanced-degree distribution among nodes and vulnerability to noisy interactions of learned representations.  SEPT~\cite{yu2021socially} proposes a socially-aware SSL framework that integrates tri-training to capture supervisory signals that are not only from self, but also from other nodes. MHCN~\cite{yu2021MHCN} proposes a multi-channel hypergraph convolutional network to enhance social recommendation by leveraging high-order user relations. In~\cite{coldstart2020sslRec},  authors propose a pre-training task to simulate the cold-start scenarios from the users/items with sufficient interactions and take the embedding reconstruction as the pretext task, improving the quality of the learned representations. 

Despite their excellent performance, they ignore the heterogeneous behavior patterns of users in the scenario of social recommendation. In this work, the proposed framework aims to model this kind of heterogeneity to assist in further boosting recommendations' performance.

\subsection{Datasets} 
We conduct experiments on two real-world datasets: Dianping~\cite{dianping2015RecSys} and Ciao~\cite{tang2012dataset}. Since we aim at producing Top-K recommendations, we leave out the ratings less than 4 and utilize the rest in these two datasets, where users' ratings for items range from 1 to 5.
The statistics of the datasets are shown in Table~\ref{tab:dataset_statistics}. For each dataset, the ratio of splitting the interactions into training, validation, and testing set is $8:1:1$.  

 \subsection{Baselines and Implementation}
 We conduct experiments by comparing with various kinds of baselines, including MF-based (BPR~\cite{rendle2009bpr}, SBPR~\cite{cikm2014sbpr}, SoRec~\cite{cikm2008SoRec}, SocialMF~\cite{recSys2010SocialMF}), GNNs-based (Diffnet~\cite{wu2019diffNet}, NGCF~\cite{wang2019ngcf}, and LightGCN~\cite{he2020lightgcn}), as well as SSL-enhanced recommendation methods (SGL~\cite{he2021sgl}, MHCN~\cite{yu2021MHCN} and SEPT~\cite{yu2021socially}). 
The baselines LightGCN+Social and SGL+Social are implemented by adding social information into their adjacency matrices for propagation via GNNs techniques in social recommendation tasks.  For implementation, BPR, SBPR, SoRec, SocialMF, Diffnet, MHCN, and SEPT are from the open-source library QRec.
The rest baselines are implemented based on the implementation of LightGCN~\cite{he2020lightgcn}. 
For a fair comparison, we apply grid search to fine-tune the hyperparameters of the baselines, initialized with the optimal parameters reported in the original papers. We use Adam optimizer to optimize all these models. 

\subsection{Evaluation Metrics}
For the metrics, we follow the ranking protocol used in ~\cite{he2021sgl} to evaluate the top-K recommendation performance and report the average \textit{NDCG}, \textit{Recall}, and \textit{Precision}, where \textit{K}=5. Higher values of these three metrics indicate better predictive performance.

\begin{table}[b]
\centering
\caption{Dataset Statistics}
\label{tab:dataset_statistics}
\scalebox{0.95}{
\begin{tabular}{c|ccccc}
\hline
Dataset  & \#Users & \#Items & \#Ratings & \#Relations & Density \\ \hline\hline
Dianping & 16,396   & 14,546   & 51,946          & 95,010       & 0.022\% \\
Ciao     & 7,375    & 105,114  & 284,086         & 111,781      & 0.037\% \\
\hline
\end{tabular}
}
\end{table}

\end{document}